# A SIMULATION OF THE PHOTOIONIZATION OF H- TOGETHER WITH THE SUBSEQUENT TRACKING OF THE LIBERATED ELECTRONS*

R. Thurman-Keup†, M. El Baz, V. Scarpine, Fermi National Accelerator Laboratory, Batavia, USA

*Abstract*

The Proton Improvement Plan - II (PIP-II) is a new linear accelerator (LINAC) complex being built at Fermilab. It is based on superconducting radiofrequency cavities and will accelerate H- ions to 800 MeV kinetic energy before injection into the existing Booster ring. Measurements of the profile of the beam along the LINAC must be done by non-intercepting methods due to the superconducting cavities. The method chosen is photoionization of a small number of H- by a focused infrared laser, aka laserwire. The number of ionized electrons is measured as a function of laser position within the H- beam. To aid in the design of the collection mechanism, a simulation was written in MATLAB with input from the commercial electromagnetic simulation, CST. This simulation calculates the number and positions of the liberated electrons and tracks them through the magnetic collection and H- beam fields to the collection point. Results from this simulation for various points along the LINAC will be shown.

## INTRODUCTION

Fermilab is in the process of constructing a new superconducting (SC) linear accelerator to replace the existing normal conducting LINAC. This project is called the Proton Improvement Plan - II (PIP-II) [1] and is being built to increase the deliverable beam intensity to the Deep Underground Neutrino Experiment being constructed in South Dakota. Since the LINAC is mostly superconducting, the use of physical wire scanners as beam profilers is not allowed in much of the LINAC due to the risk of a broken wire contaminating the SC cavities. As the beam is composed of H- ions, laserwires [2,3] were chosen as the profiler. A laserwire functions by photoionization of the extra electron on the H- ion using a focused laser. The binding energy of the extra electron is only 0.7542 eV and can be ionized by a 1064 nm YAG laser. The ionized electrons, which are proportional in number to the local density of H- ions, are collected as the laser is moved through the H- beam, enabling reconstruction of the profile of the beam. Due to the complexity of the system, a simulation was developed to aid in the planning of the laser optics and the electron collection and is presented in this paper.

## EXPERIMENTAL DEVICE

The laserwire system for PIP-II is comprised of a source laser that is transported through a pipe from the laser room to the H- beamline (Fig. 1). The pipe is under vacuum to both reduce distortions from air currents and to serve as a safety interlock system. At each measurement location (station), there is an insertable mirror to direct the laser down to the beamline.

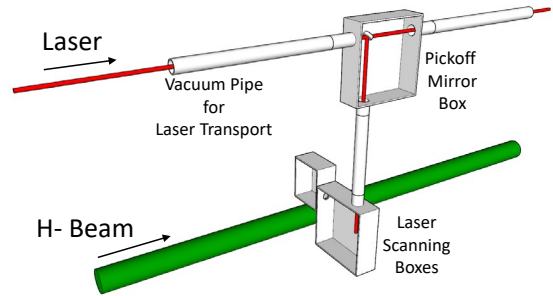

Figure 1: Diagram of laser transport. The laser originates in the laser room and is transported through a vacuum line to the end of the LINAC. At each laserwire, a vertical pipe can direct the laser to the H- beamline.

The optical system at the beamline is comprised of movable stages to scan the laser across the H- beam and to select vertical or horizontal scan mode (Fig. 2).

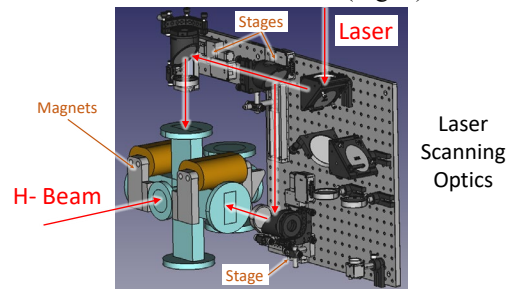

Figure 2: Optics inside the laser scanning box. There are linear stages to pick horizontal or vertical scan, and to do the scans.

There are 12 beamline stations with the locations and beam parameters summarized in Table 1 (plus an additional emittance measuring one after the LINAC). The laserwires are installed just downstream of the location column entry which specifies the cryomodule section and the cryomodule number within that section. The optics within the beamline stations focus the laser from a large incoming rms size of ~5 mm to a focused rms size of <100 μm. The focusing distance will be approximately 300 mm. Once the laser has ionized the H- beam, the electrons are bent vertically upward to a detector which will nominally be a faraday cup (Fig. 3).

## SIMULATION

The simulation is comprised of three parts: the calculation of the number of photoionizations and generation of the electrons, the calculation of the electromagnetic fields of the beam and the magnet, and the tracking of the electrons to the detector.

Table 1: Laserwire Locations and Expected Beam Parameters

| Laserwire Location | Position [m] | Beam Parameters | | | |
|---|---|---|---|---|---|
| | | $E_k$ [MeV] | $\sigma_x$ [mm] | $\sigma_y$ [mm] | $\sigma_t$ [ps] |
| MEBT | 18.7 | 2.1 | 2.3 | 2.3 | 208 |
| HWR | 25.4 | 10.0 | 1.3 | 1.4 | 33 |
| SSR1 CM #1 | 31.6 | 18.6 | 1.2 | 1.2 | 20 |
| SSR2 CM #2 | 51.5 | 61.2 | 1.4 | 1.5 | 13 |
| SSR2 CM #4 | 65.2 | 106.0 | 1.3 | 1.3 | 13 |
| SSR2 CM #6 | 78.9 | 153.9 | 1.4 | 1.4 | 11 |
| LB650 CM #1 | 93.4 | 192.2 | 2.3 | 2.4 | 6.7 |
| LB650 CM #3 | 107.1 | 267.4 | 1.8 | 1.9 | 5.7 |
| LB650 CM #6 | 127.6 | 400.9 | 1.9 | 2.0 | 5.2 |
| LB650 CM #9 | 148.0 | 516.5 | 2.1 | 1.8 | 5.0 |
| HB650 CM #2 | 170.5 | 652.6 | 1.8 | 1.9 | 4.4 |
| HB650 CM #4 | 192.9 | 833.3 | 1.5 | 1.9 | 3.7 |

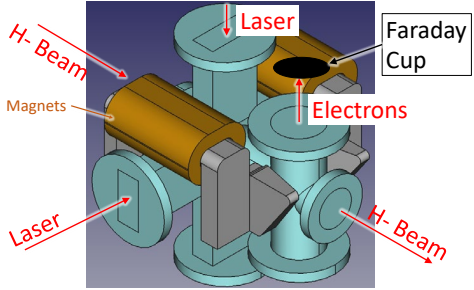

Figure 3: The incoming H- beam is partially ionized by the laser and the electrons are bent vertically by a magnet to a faraday cup. The blue volume is the vacuum chamber.

*Photoionization Simulation*

Calculation of the number of photoionizations requires the particle densities of both the incoming laser and H- beams, and the cross section for photoionization

$$dn = c\sigma_I \sqrt{\left(\left|\hat{l} - \vec{\beta}\right|^2 - \left|\hat{l} \times \vec{\beta}\right|^2\right)} I_l I_b \, dt dx dy dz \quad (1)$$

Here $c$ is the speed of light, $\hat{l}$ is the laser beam direction, $\vec{\beta}$ is the H- beam velocity vector divided by $c$, the square root is a Lorentz-transformed relative velocity term [4], $I_l$ and $I_b$ are the laser and H- number densities respectively, and $\sigma_I$ is the photoionization cross section in the frame of the H-as a function of the laser wavelength Lorentz-transformed to the frame of the H-

$$\tilde{\lambda}_l = \lambda_l \frac{\sqrt{1 - \beta^2}}{1 - \hat{l} \cdot \vec{\beta}} \quad (2)$$

where $\lambda_l$ is the laser wavelength in the lab frame.

For the PIP-II laserwire design, the laser is transverse to the H- beam and as such, the square root term in Eq. 1 and the denominator in Eq. 2 are both equal to 1. The Lorentz-transformed laser wavelength shifts from 1064 nm in the lab frame to 1062 nm in the MEBT and to 883 nm at the end of the LINAC. The cross sections, $\sigma_I$, are extracted at the desired wavelengths from a polynomial fit to Table IV in reference [5].

The simulation code is written in MATLAB [6]. It steps through time and calculates the number of ionized electrons on a predefined set of physical mesh points. The simulation utilizes one of two possible numerical mesh arrangements, laser grid or beam grid, depending on the laser and H- beam parameters.

The laser grid approach creates a fixed cylindrical mesh axially aligned with the laser and covering the overlap region of the laser and the H- beam (Fig. 4). The mesh is an elliptic cylinder with radii that scale with the transverse laser size. This keeps the mesh size proportional to the laser width and avoids losing resolution near the laser waist. This works well when the laser intensity is low enough to avoid significant depletion of the H- beam which is not accounted for in this approach.

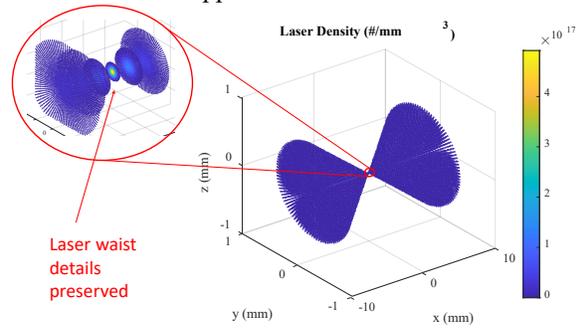

Figure 4: Laser grid version of the mesh points. The mesh spacing scales with the diameter of the laser as it propagates though the H- beam.

The beam grid version is a rectangular fixed mesh covering the physical overlap region of the laser and the H- beam. The simulated time steps have a spacing that is equal to the physical mesh spacing divided by the H- beam velocity. This relationship means that with every time step, the H- beam moves one spatial step, allowing for the fixed mesh to handle depletion of the H- (Fig. 5).

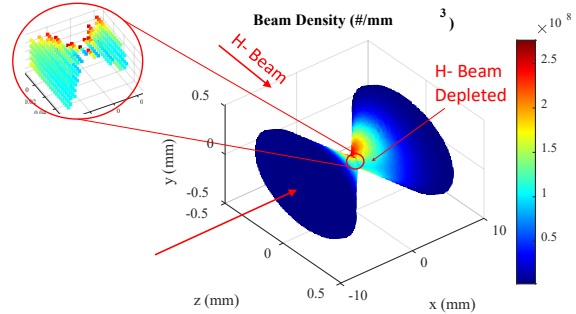

Figure 5: Beam grid version of the mesh points showing the H- beam intensity. Note the depletion of the H- beam as it passes through the laser.

In both cases, the mesh extent is determined from the space-time interaction region of the laser and H- while the desired number of mesh points is set by the user to avoid loss of details such as the laser waist.

Electrons for tracking are generated from the space-time mesh containing the calculated number of photoionizations. A Poisson distribution is utilized for up to 500 electrons, beyond which a Gaussian approximation is used. The electrons are given momenta based on the 6D phase space specified for the H- beam (Fig. 6).

*Electromagnetic Field Calculations*

The electromagnetic field calculations involve two separate parts: the fields of the H- beam, and the static fields of both the electron collection magnets and possibly secondary electron containment electrodes.

The beam field calculation is done in MATLAB and uses a single bunch which can have any arbitrary shape and size, but a single fixed velocity. For PIP-II, the H- beam is bunched at a frequency of 162.5 MHz and we use a gaussian shape in all three dimensions. The fields are evaluated on a rectangular mesh containing an inner section with generally closer spacing to resolve the fields within the bunch, and an outer section with larger spacing. The fields at each mesh point are a numerical integration over the bunch charge. This numerical integration requires a mesh of the bunch as well. To reduce numerical errors, the field evaluation mesh spacing is adjusted such that it is an integer multiple of the bunch mesh spacing. The results of the field evaluation are stored for later use by the tracking simulation.

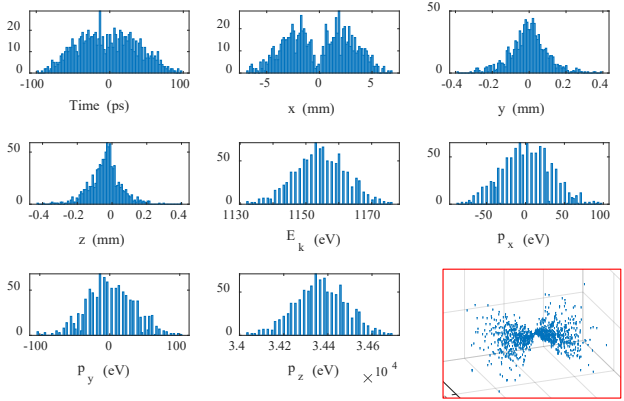

Figure 6: Generated electron distributions. The laser waist can be seen in the $x$ position distribution. A 3D distribution of the electron positions is shown in the red box where the laser waist can be seen again.

The static fields are calculated in CST [7] which is a 3D electromagnetic simulation. We designed a magnet with a return yoke to reduce the impact on the H- beam (Fig. 7). In addition, we will also add a small quadrupole magnet for the MEBT laserwire to compensate electron spreading from H- space charge forces. When the electrons strike the faraday cup, they generate secondary electrons that may escape the surface of the faraday cup, altering the collected charge. To avoid this, we may need to install a conductive ring to apply an electrostatic field to keep the secondary electrons from leaving. As with the bunch fields, these electrostatic results are also stored for later use.

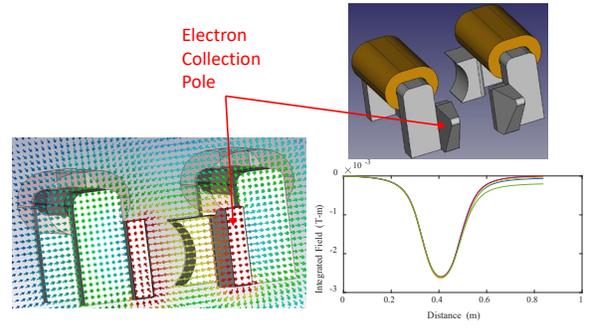

Figure 7: CST calculation of magnetic field of magnets for electron collection. The plot shows transverse magnetic field integrated along H- beam direction for 5 different transverse locations.

*Tracking Simulation*

The tracking simulation is written in MATLAB and was created originally to track electrons in the electron beam profiler [8] and ionization profile monitors [9]. It tracks the electrons through static fields and the fields of the H- bunches but does not do self-interactions with the other electrons since the effect is generally much smaller than the H- bunches. It uses an adaptive Runge-Kutta method [10] to solve the pseudo-relativistic second-order differential equation of motion

$$\vec{F}(\vec{r},t) = \frac{d\vec{p}}{dt} = m\frac{d(\gamma\vec{v})}{dt}$$
$$\vec{F}(\vec{r},t) = m\gamma\left(\vec{a} + \gamma^2\vec{\beta}(\vec{\beta}\cdot\vec{a})\right) \quad (3)$$

which, when inverted to find $\vec{a}$, is

$$\vec{a} = \frac{d^2\vec{r}}{dt^2} = \frac{1}{\gamma m}\left(\boldsymbol{I} - \vec{\beta}\vec{\beta}^T\right)\vec{F}(\vec{r},t) \quad (4)$$

where $\vec{F}(\vec{r},t) = q\left(\vec{E}(\vec{r},t) + \vec{v}\times\vec{B}(\vec{r},t)\right)$, $m$ is the mass of the particle being tracked, $\gamma = 1/\sqrt{1-\beta^2}$ is the Lorentz factor, and $\boldsymbol{I}$ is the identity matrix. We apply the Runge-Kutta method to the second-order differential equation rewritten as coupled first order differential equations

$$\frac{d\vec{v}}{dt} = \frac{1}{\gamma m}\left(\boldsymbol{I} - \vec{\beta}\vec{\beta}^T\right)\vec{F}(\vec{r},t)$$
$$\frac{d\vec{r}}{dt} = \vec{v} \quad (5)$$

For each Runge-Kutta time step, the previously stored bunch and static fields are interpolated to find the value at the space-time location of the particle being tracked. For the bunch fields, the field is interpolated to the requested position after adjusting it for the requested time and the velocity of the bunch,

$$\vec{E}|\vec{B}(\vec{r},t) \rightarrow \vec{E}|\vec{B}(\vec{r} + \vec{v}t_m) \quad (6)$$

where $t_m = (t \bmod t_b)$. The modulo function implements a repetitive bunch structure at the specified bunch spacing, $t_b$.

The adaptive part of the algorithm adjusts the step size to keep changes in the momenta, either absolute value or direction, within a range specified by thresholds. If any of the thresholds are exceeded, the step size is adjusted to compensate.

## RESULTS

The simulation has been used to design the collection magnet and find optimal laser parameters to avoid wasted laser energy and resolution degradation. Figures 8 and 9 show electron trajectories at two locations: MEBT and HWR. The MEBT location (Fig. 8) shows tracking results with and without a quadrupole field. The extra quadrupole may be necessary in the MEBT to deal with spreading of the electrons due to the space charge of the H- beam.

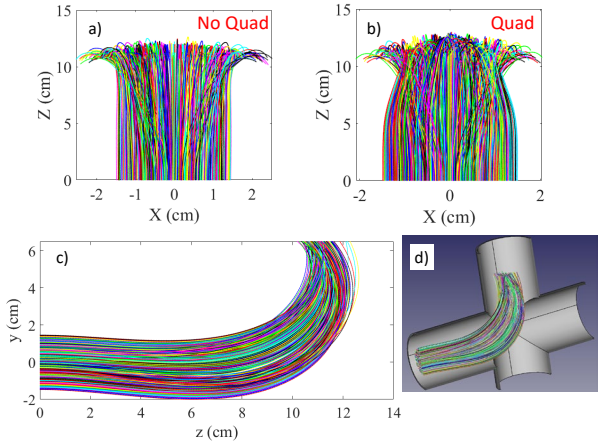

Figure 8: MEBT laserwire; a) and b) are the trajectories looking down from above the beam. They show the difference in final spread between having a quadrupole field and not having a quadrupole field. Plots c) and d) are the trajectories looking from the side of the beam. They first deviate down before bending up which allows more clearance between the electrons and the beampipe.

Figure 9 shows trajectories in the laserwire after the HWR which is the first cryomodule.

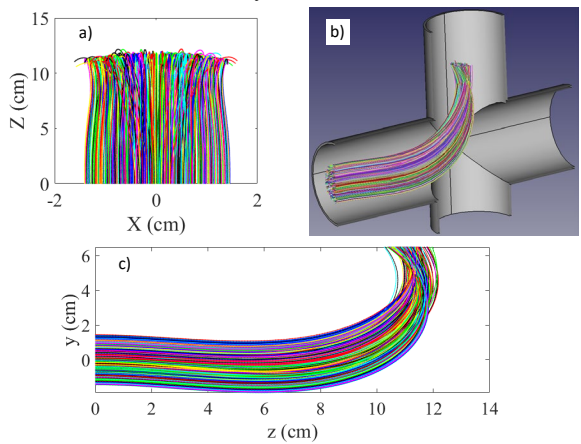

Figure 9: HWR laserwire. Plot a) is the trajectories looking down from above the beam. b) and c) are the trajectories looking from the side of the beam.

The containment of the electrons is better at this location since the electrons have higher energy and are less susceptible to H- space charge forces. The initial deflection down in all of these is driven by the positioning of the two magnet yokes where the first corrector yoke is closer to the laser interaction region. This effect helps to increase the clearance between the electrons and the corner of the beampipes.

Studies were also done to evaluate the photoionization rates for different length laser pulses, laser arrival time jitter, and H- temporal bunch lengths (Fig. 10). These results will help to determine the optimal laser properties to maximize signal and resolution while minimizing unused laser energy which has detrimental effects on the vacuum windows. For instance, from the plots, we can see that laser pulses with a length of 20 or 30 ps have good photoionization but small variation with jitter.

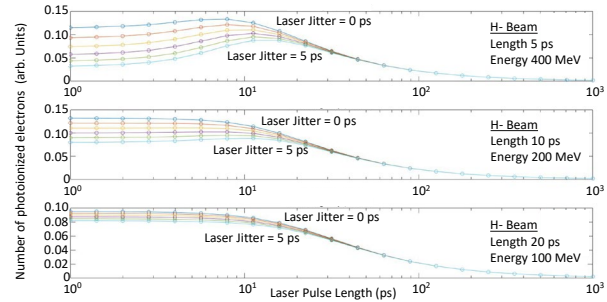

Figure 10: Study of the effect of laser jitter on the number of photoionized electrons using the simulation at 3 different H- beam energies.

This simulation was also recently used in an analysis of a novel bunch length measurement at the Spallation Neutron Source at Oak Ridge National Lab [11].

## CONCLUSION

This laserwire simulation has proven to be useful for a number of analyses pertaining to laserwires. From magnet designs and laser parameter choices to measurement analyses. We foresee even more uses as we proceed to build and commission the PIP-II laserwire systems.